\documentclass[review]{elsarticle}

\usepackage{lineno,hyperref}
\modulolinenumbers[5]

 	\usepackage{algorithm2e}
  	\usepackage{graphics}
  	\usepackage{footnote}
  	\usepackage{amssymb}
  	\usepackage{amsthm}
  	\usepackage{amsmath}
  	\usepackage{amsfonts}
  	\usepackage{amsxtra}
  	\usepackage{color}
  	\usepackage{cite}
 	\usepackage{colortbl}
 	\usepackage{booktabs}
	\usepackage[caption=false]{subfig}
    \usepackage{graphicx}
	\usepackage{multirow}
    \usepackage{multicol}
    \usepackage[utf8]{inputenc}
    \usepackage{dblfloatfix}
    \usepackage{float}
    \usepackage{filecontents}
    \usepackage{subfig}
    \usepackage{mathtools}
	\usepackage{stackengine}
	\usepackage[nodots]{numcompress}
    \usepackage{natbib}
    \usepackage{comment}

\journal{arxiv.org}

\usepackage{silence}
\usepackage{silence}
\usepackage{silence}
\usepackage{numcompress}\biboptions{authoryear}

\begin{document}

\begin{frontmatter}

\title{Classification of glomerular hypercellularity using convolutional features and support vector machine}

%% Group authors per affiliation:
  \author[add1]{Paulo Chagas\corref{cor1}}
  % \ead{paulo.chagas@ufba.br}
   \author[add1]{Luiz Souza\corref{cor1}}
 %  \ead{luiz.otavio@ufba.br}
   \author[add2]{Ikaro Araújo}
   %\ead{ikaroa@ufba.br}
   \author[add3]{Nayze Aldeman}
  %\ead{nayzealdeman@gmail.com}
   \author[add4]{Angelo Duarte}
   %\ead{angeloduarte@uefs.br}
   \author[add4]{Michele Angelo}
  %\ead{mfangelo@ecomp.uefs.br}
     \author[add5]{Washington LC dos-Santos\corref{cor2}}
     %\ead{wluis@bahia.fiocruz.br}
      \author[add1]{Luciano Oliveira\corref{cor2}}
    %\ead{lrebouca@ufba.br}
  
   \cortext[cor1]{First two authors contributed equally.}
   \cortext[cor2]{Corresponding authors: Luciano Oliveira, email: lrebouca@ufba.br, Washington L-C dos Santos, email: wluis@bahia.fiocruz.br}
   
   \address[add1]{IVISION Lab, Universidade Federal da Bahia, Bahia, Brazil}
   \address[add2]{PPGM, Universidade Federal da Bahia, Bahia, Brazil}
   \address[add3]{Departamento de Medicina Especializada - Universidade Federal do Piauí, Piauí, Brazil}
   \address[add4]{Universidade Estadual de Feira de Santana, Bahia, Brazil}
   \address[add5]{Fundação Oswaldo Cruz - Instituto Gonçalo Moniz, Bahia, Brazil}

\begin{abstract}
Glomeruli are histological structures of the kidney cortex formed by interwoven blood capillaries, and are responsible for blood filtration. Glomerular lesions impair kidney filtration capability, leading to protein loss and metabolic waste retention. An example of lesion is the glomerular hypercellularity, which is characterized by an increase in the number of cell nuclei in different areas of the glomeruli. Glomerular hypercellularity is a frequent lesion present in different kidney diseases. Automatic detection of glomerular hypercellularity would accelerate the screening of scanned histological slides for the lesion, enhancing clinical diagnosis. Having this in mind, we propose a new approach for classification of hypercellularity in human kidney images. Our proposed method introduces a novel architecture of  a convolutional neural network (CNN) along with a support vector machine, achieving near perfect average results with the FIOCRUZ data set in a binary classification (lesion or normal). Our deep-based classifier outperformed the state-of-the-art results on the same data set. Additionally, classification of hypercellularity sub-lesions was also performed, considering mesangial, endocapilar and both lesions; in this multi-classification task, our proposed method just failed in 4\% of the cases. To the best of our knowledge, this is the first study on deep learning over a data set of glomerular hypercellularity images of human kidney.
\end{abstract}


\end{frontmatter}

%\linenumbers

\section{Introduction}

Digital histopathology is a research field that exploits digital images for the analysis of tissue samples. The digital pictures are obtained either by scanning histological whole-slide-images (WSIs) or by collecting snapshots of histological structures relevant for the diagnosis of diseases \citep{al2012digital}. This approach makes gathering large-scale data sets of histological lesions easier to review or to exchange information among pathologists without the inconvenience of working with the actual glass slides. The evolution of the computer vision field impacted the entire digital medicine, supporting pathologists on the automatic analysis of various types of medical images, as well as improving the accuracy of computer-aided diagnosis \citep{MADABHUSHI2016170,LITJENS201760}.

In the special case of renal histopathology, disease markers are mostly found in the glomeruli, presenting highly diverse and heterogeneous characteristics. The glomerulus is a histological structure from the kidney cortex, formed by a network of capillaries charged of performing blood filtration. As an elementary filtering structure, it is targeted with many primary and systemic diseases, leading to different patterns of glomerular lesions. Finding and classifying glomerular lesions are fundamental steps toward the diagnosis of many kidney diseases. These tasks rely on the expertise of pathologists and much effort has been made to better define and create consensus about relevant lesions. In fact, after successive discussion and validation studies in the field, increased consistency has been achieved in the diagnosis and classification of glomerular renal diseases such as lupus nephritis, IgA nephropathy, and rejection of kidney transplant \citep{bajema2018revision,trimarchi2017oxford,JOOSTEN20051}. Some limiting factors to the performance of histological diagnosis are the complexity of lesions, which, in some cases, may impair a clear definition in terms of criteria and consequently a suitable agreement among pathologists \citep{barisoni2013digital}.
        
\begin{figure}[!t]
	\centering
	\includegraphics[width=0.7\textwidth]{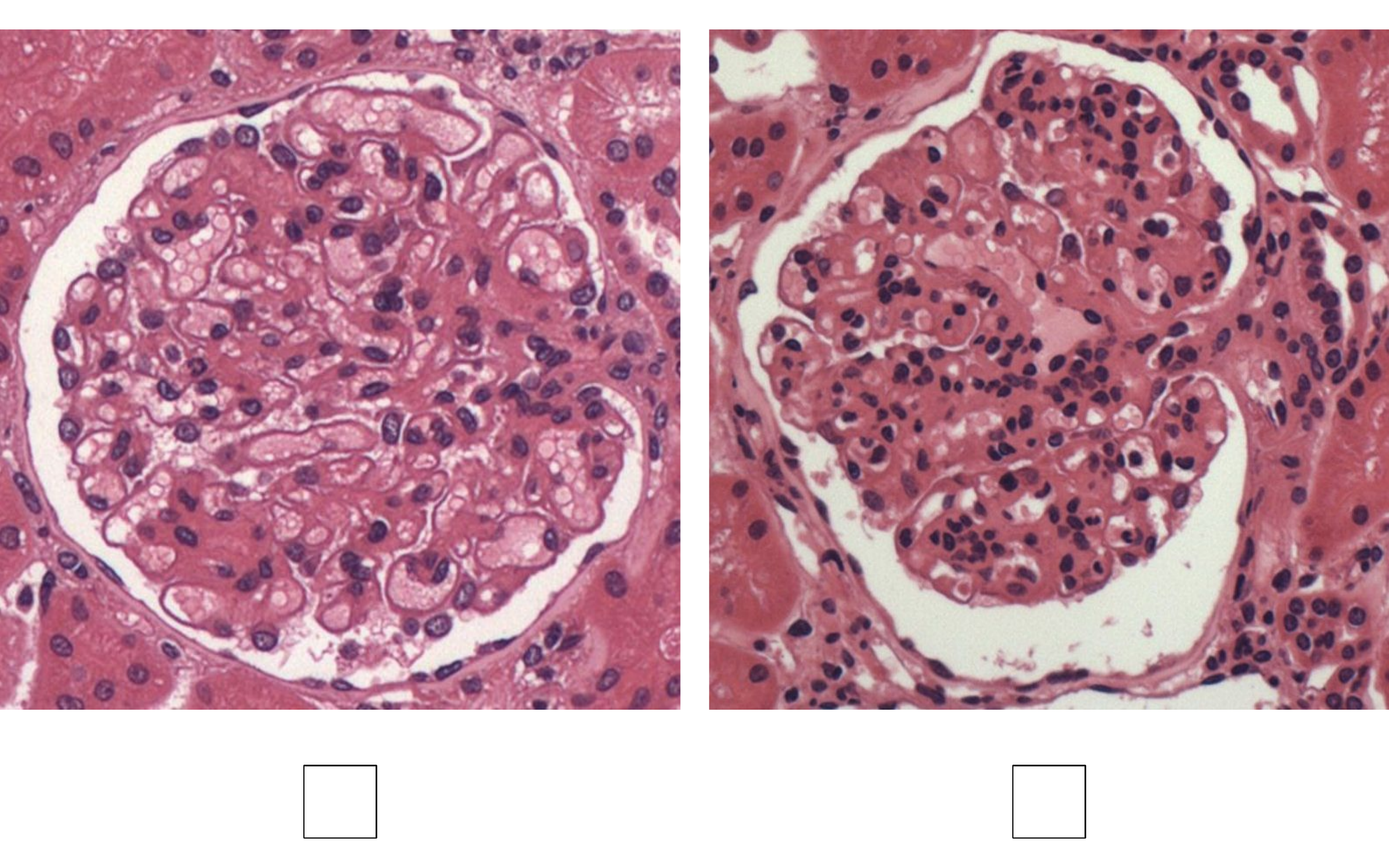}
	\caption{Mark with an $X$ on the image with hypercellularity lesion.}
    \label{fig01} 
\end{figure}

Particularly, glomerular hypercellularity is a frequent lesion found in kidney biopsies, defined by an increase in the number of cells in the glomeruli. This type of lesion is an integral component of many glomerular diseases such as proliferative and membranoproliferative glomerulonephritis, being a marker of activity in lupus and IgA nephropathy \citep{bajema2018revision,trimarchi2017oxford}. Hypercellularity can be identified by a careful look at the histological sections from the glomeruli, searching for the presence of agglomerates formed by four or more cell nuclei in the mesangial area (mesangial hypercellularity), or by cell aggregates that fill the capillary lumen (endocapillary hypercellularity) \citep{Renal_disease_1995,fogo2003approach}. Figure~\ref{fig01} shows the complexity of this problem, and the following question can be raised: \textit{Which image depicts a glomerulus with a hypercellularity lesion?} The answer to this question is \textit{the image on the right} due to the increased nuclei density; on the left, the image shows an example of a normal glomerulus with no significant number of cell clusters.

Although hypercellularity is easy to define and usually easy to be assessed in histological sections, an agreement among pathologists may decrease for focal hypercellularity and for occurrences in specific regions of the glomerulus. For instance, a recent report from the IgA \textit{Nephropathy Classification Working Group} showed inconsistencies among specialist even in the use of dichotomous MEST system scores such as E (endocapillary hypercellularity) and M (mesangial hypercellularity) \citep{trimarchi2017oxford}. Correct assessment of these scores is crucial for relevant clinical-pathological correlation and for predicting the patient outcome. A consistent glomerulus classification can be deemed as an important and difficult step towards diagnosing a renal disease in a biopsy evaluation \citep{pedraza17}. 

Some works have already approached the tasks of glomerulus identification and segmentation \citep{sarder16,KANNAN2019,simon18}, which are useful in situations that require an analysis of the entire WSI.   \citet{barros2017pathospotter} proposed a method relying on classical image pre-processing techniques and a k-nearest neighborhood to classify hypercellularity lesions; that work used 811 images of human glomeruli (referred here as FIOCRUZ data set) stained with hematoxylin-eosin (H\&E) and periodic acid–Schiff (PAS) from a set of biopsy slides. More recently, deep neural networks outperformed handcrafted features for some tasks on histological images as well, achieving stunning results in different scenarios \citep{janowczyk2016deep,xu16,sharma17,wahab17,spanhol16,hou16,zhang2018deep,FABIJANSKA20181,GANDOMKAR201814}. In particular to glomerular detection with deep-learning, \citet{marsh18} introduced a convolutional neural network for automatic localization of glomeruli, further classifying global glomerulosclerosis in donor kidney biopsies for transplantation. 

An automated process for glomerular lesion classification would have many applications, such as: Large-scale classification of cases based on histological images, consistency of morphological classification, and identification of tissue markers of disease progression.

\subsection{Contributions} 

Three main contributions are brought here: (i) Instead of using conventional classification methods as in \citep{barros2017pathospotter}, we propose a CNN-based architecture to extract trainable features to represent a glomerulus, (ii) by using the proposed CNN as a feature extractor, an SVM classifies the CNN features as a normal or a injured glomerulus, (iii) we also extend the proposed model for classification of specific hypercellularity lesions (endocapillary hypercellularity, mesangial hypercellularity, and both), providing an analysis of the generated features for both binary and multi-lesion classification. The final CNN-SVM classifier reached near perfect results in four different train/test splits of the data set introduced in \citep{barros2017pathospotter}; in the multi-classification task, the same architecture failed in just 4\% of the cases in ten-fold cross-validation study. At the end, the misclassified images were analyzed by three pathologists, showing that there were no consensus for most of those images.

\section{Classifying glomerular hypercellularity}

The classification of a glomerular hypercellularity lesion could be tackled as defining areas and counting nuclei. If the number of nuclei per area surpasses a threshold, one can diagnose a glomerulus as with a hypercellularity lesion. Instead of following this pathologist-annotation approach, an automatic classification consists of using examples of histological images to train a classifier. A histological image is a 2-dimensional grid of pixels that brings specific information such as colors, edges, shapes, textures, which can be general or specific to classify a glomerular lesion. Consequently, conceiving a successful feature extractor demands some domain expertise, which brings us to the following question: \textit{What is the best feature set for classifying glomerular hypercellularity lesions?}

Many feature extraction techniques are available in the literature, and a specific method could be designed as well. In contrast to conventional classifiers, deep-learning aims to automatically learn hierarchical feature representations of the input data, without the need of creating any particular feature extractor \citep{lecun2015deep}. Our work proposes a novel CNN-based architecture for glomerular hypercellularity classification. After training a CNN, it is possible to use a strong classifier on the convolutional backbone of features. This way, we propose to use a CNN architecture to extract trainable features, which ultimately will feed an SVM to carry on the final classification. The proposed architecture is evaluated for both binary and multi-class classification. The rationale to use an SVM is based on the main characteristic of this classifier that is to cast optimization problems, which are convex and quadratic. Ultimately, these characteristics guarantee that the hyperplane found is the optimum one. The second reason is to analyze the behavior of feature space extracted from the CNN, which empirically demonstrated to be linear, in our experiments. Linearity in the feature space is expected to provide faster and higher results.  

\subsection{Conceiving the proposed CNN architecture}\label{cnn_arch}

There are several well-established CNN architectures available in the literature \citep{canziani2016analysis}, which were designed to be robust to deal with hundreds of different classes. However, these models tend to overfitting, when trained using few data. Since the data set we used \citep{barros2017pathospotter} consists of a small training set, we decided to build our own architecture from scratch, modifying it accordingly to our needs. The ultimate goal is to focus on achieving a high accuracy, avoiding overfitting.
		
A CNN architecture is organized in layers, each one applying a specific operation. Although there are many variations of CNN architectures, they share some basic components, such as convolutional, pooling, and fully-connected layers \citep{gu2015recent}. The convolutional layer is the fundamental building block of a CNN model, which is comprised of various learnable kernels (filters) followed by a nonlinear activation function. A pooling layer (usually applied after a convolutional layer) is used to compute feature maps condensed in a smaller representation with the goal of achieving some invariance. After some convolutional and pooling operations, the top of the network results in a high-level representation of the input image, which is more robust than the raw pixel information, or hopefully than handcrafted features. This type of architecture requires a fully-connected layer to perform high-level classification using those features, working as a multilayer perceptron (MLP) on top of a CNN backbone.

\begin{figure}[!t]
  \centering{\subfloat[][Architecture 1] {\includegraphics[height=0.215\textwidth]{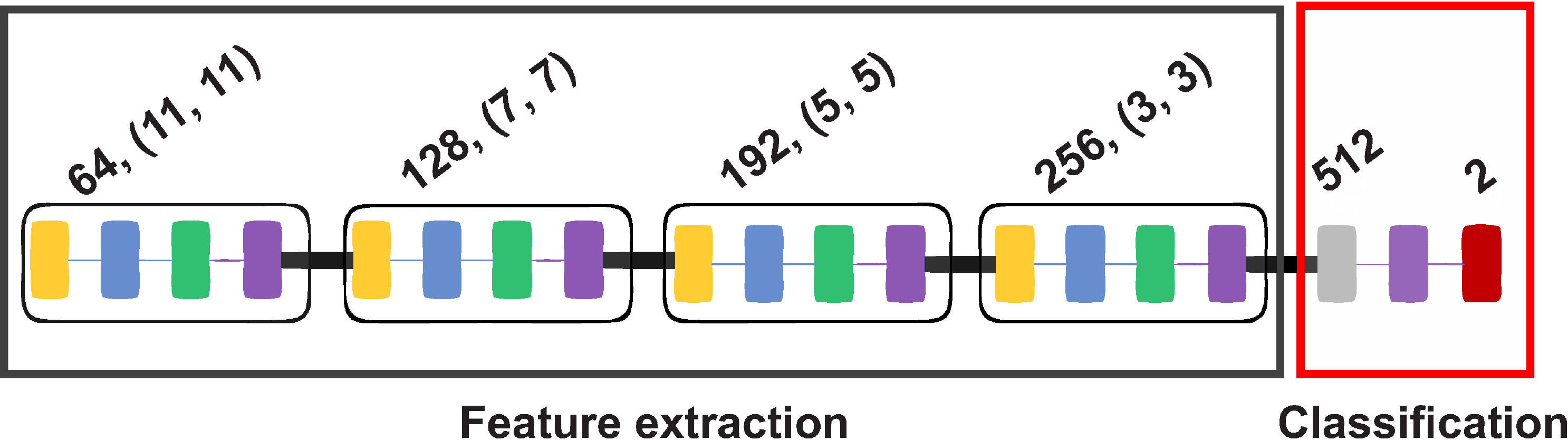} \label{figx:a}}}
  \hfil
   \centering{\subfloat[][Architecture 2]{\includegraphics[height=0.215\textwidth]{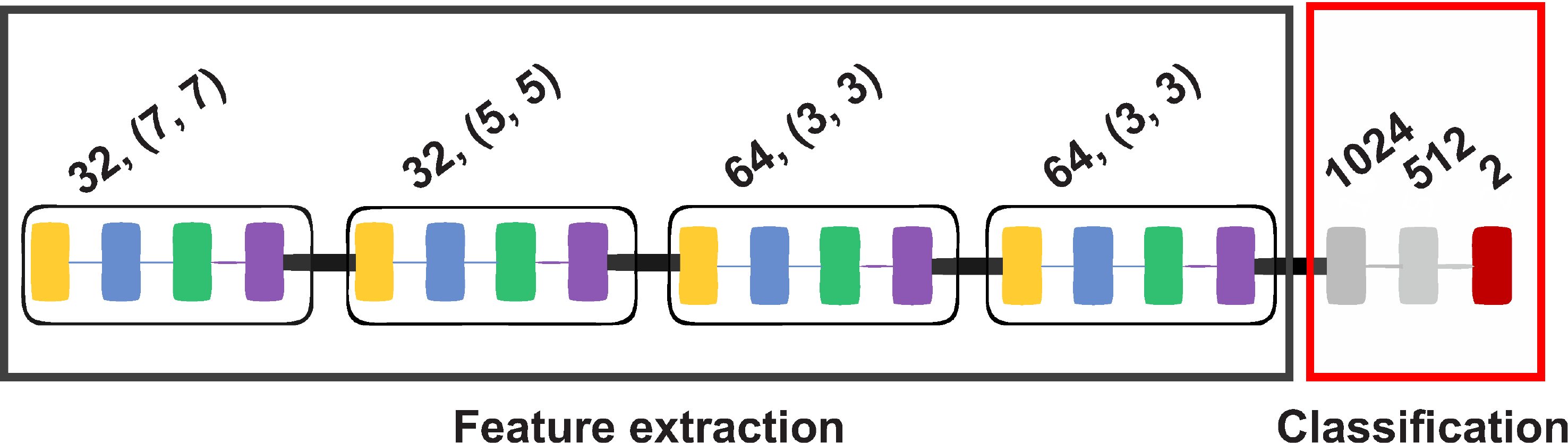} \label{figx:b}}}
  \hfil
  \centering{\subfloat[][Architecture 3]{\includegraphics[height=0.215\textwidth]{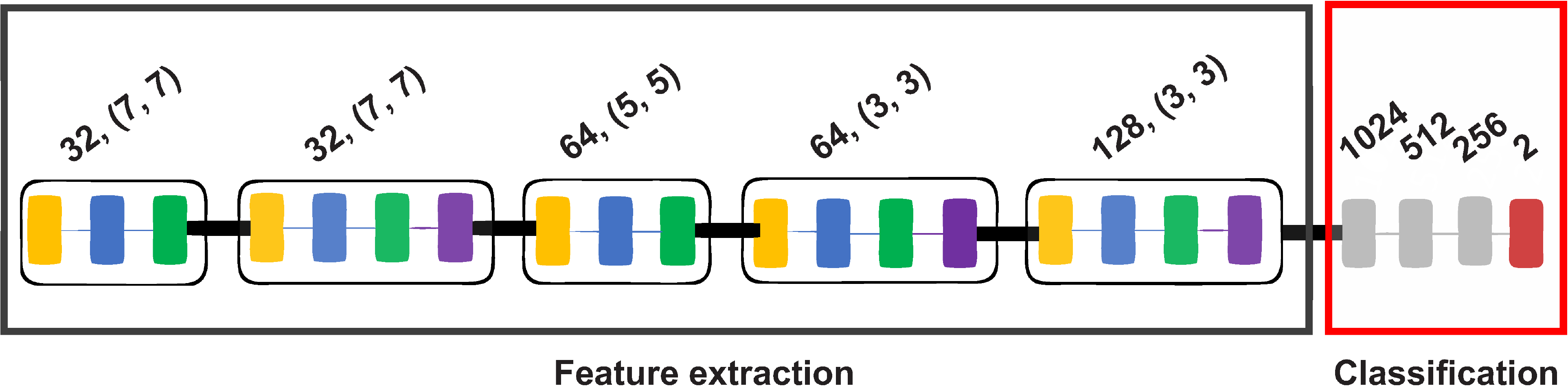} \label{figx:c}}}
   \centering{\subfloat[][Architecture 4]{\includegraphics[height=0.215\textwidth]{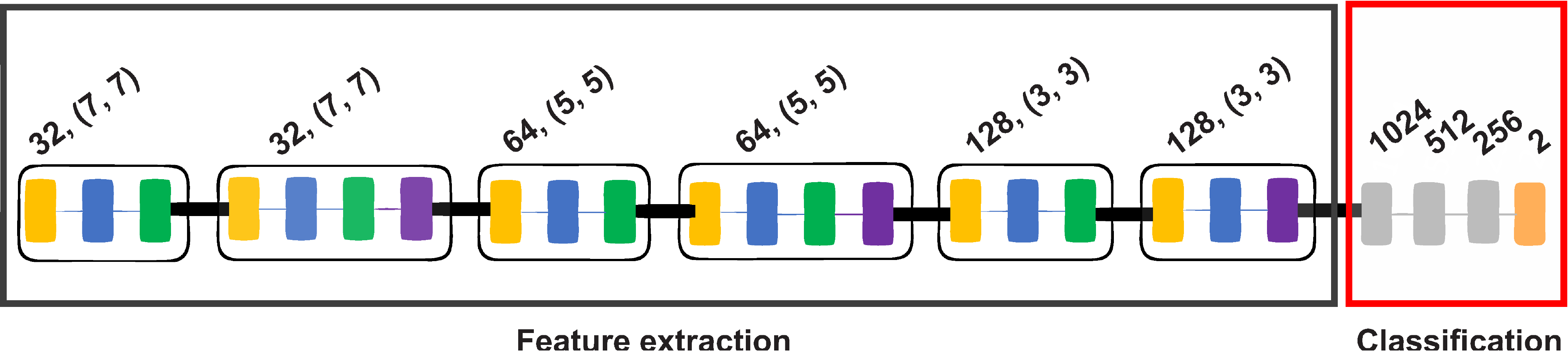} \label{figx:d}}}
  
  \bigskip
   \centering{{\includegraphics[height=0.09\textwidth]{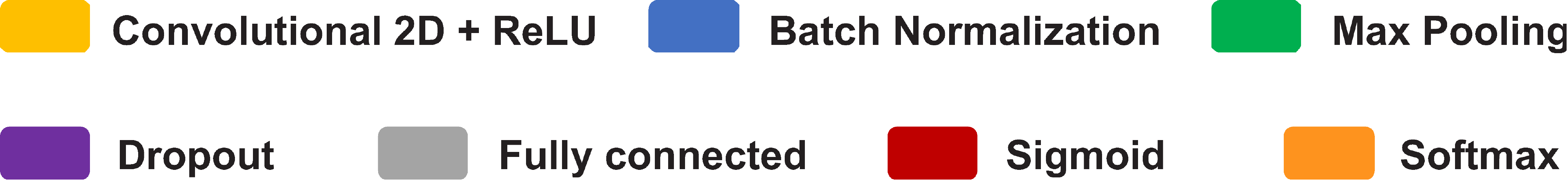}}}
 \medskip
 
 \caption{Four CNN architectures proposed here: (a) Architecture 1 and (b) architecture 2, with four convolutional layers in the backbone; (c) architecture 3 and (d) architecture 4, with five and six convolutional layers, respectively, in the backbone. } \label{arch}
  \end{figure}

Four architectures were initially implemented and Figure ~\ref{arch} highlights the convolutional blocks (CNN backbone) used for feature extraction, and the MLP blocks (fully-connected layers and final activation) used for classification. The first architecture was designed in the view of investigating how the lesion classification behaved using fewer layers. In addition to the operations previously cited, batch normalization, regularization, and dropout operations were applied to reduce overfitting. The first architecture (Fig. \ref{figx:a}) is composed mainly of four convolutional layers, with the other operations applied between those layers, followed by one fully-connected layer. A rectified linear unit (ReLu) was used as an activation function and max-function for pooling operations. For the calculation of the class probabilities after the fully-connected layers, a sigmoid function was first tried, and further changed to a soft-max function. With this first architecture in mind, updates were performed based on the stability of the accuracy curve in the validation set, and other three architectures were proposed (Figs. ~\ref{figx:b}, \ref{figx:c} and \ref{figx:d}).

\begin{figure}[t]
\centering
	\begin{multicols}{2}
		\includegraphics[width=0.45\textwidth]{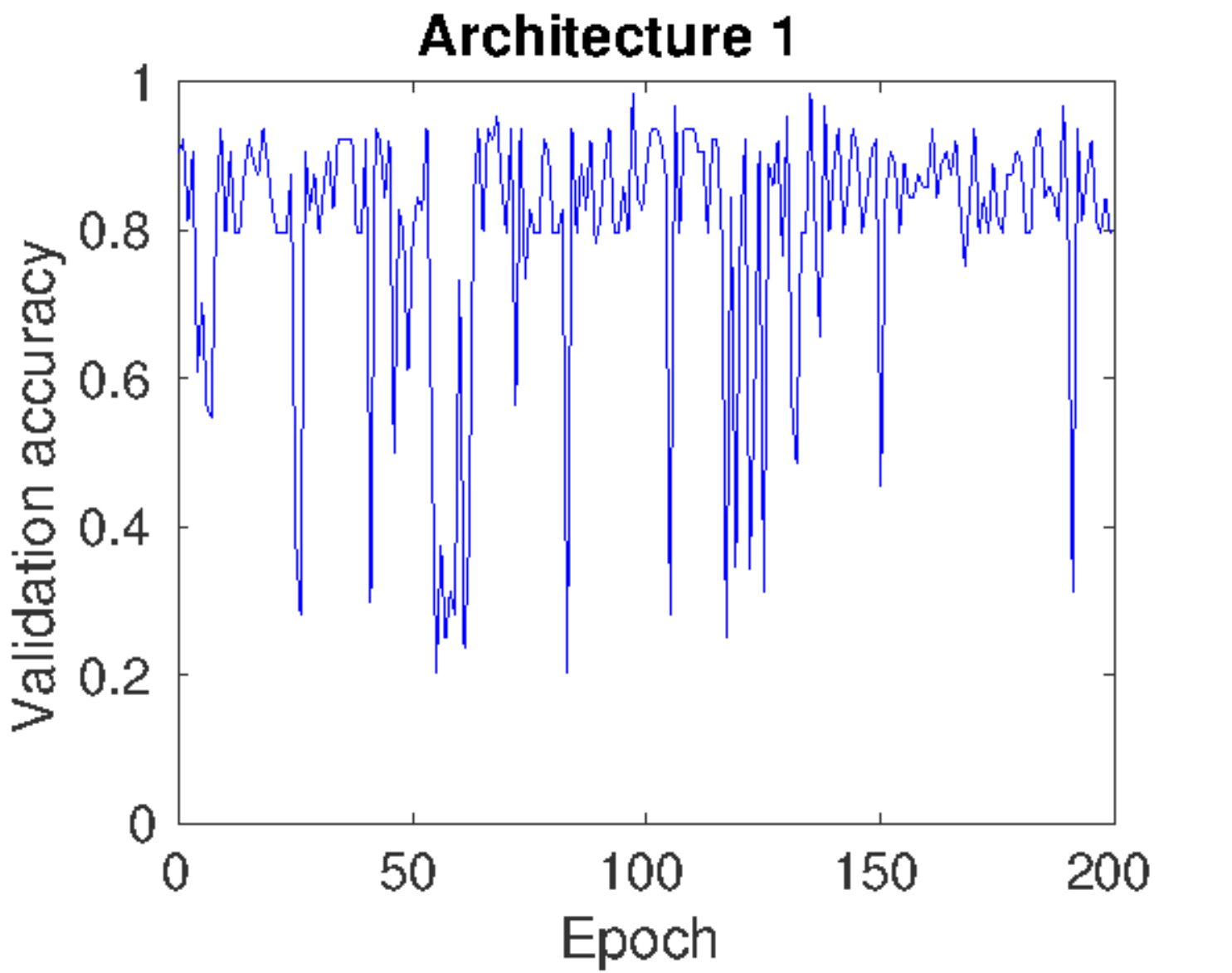}\par 
		\includegraphics[width=0.45\textwidth]{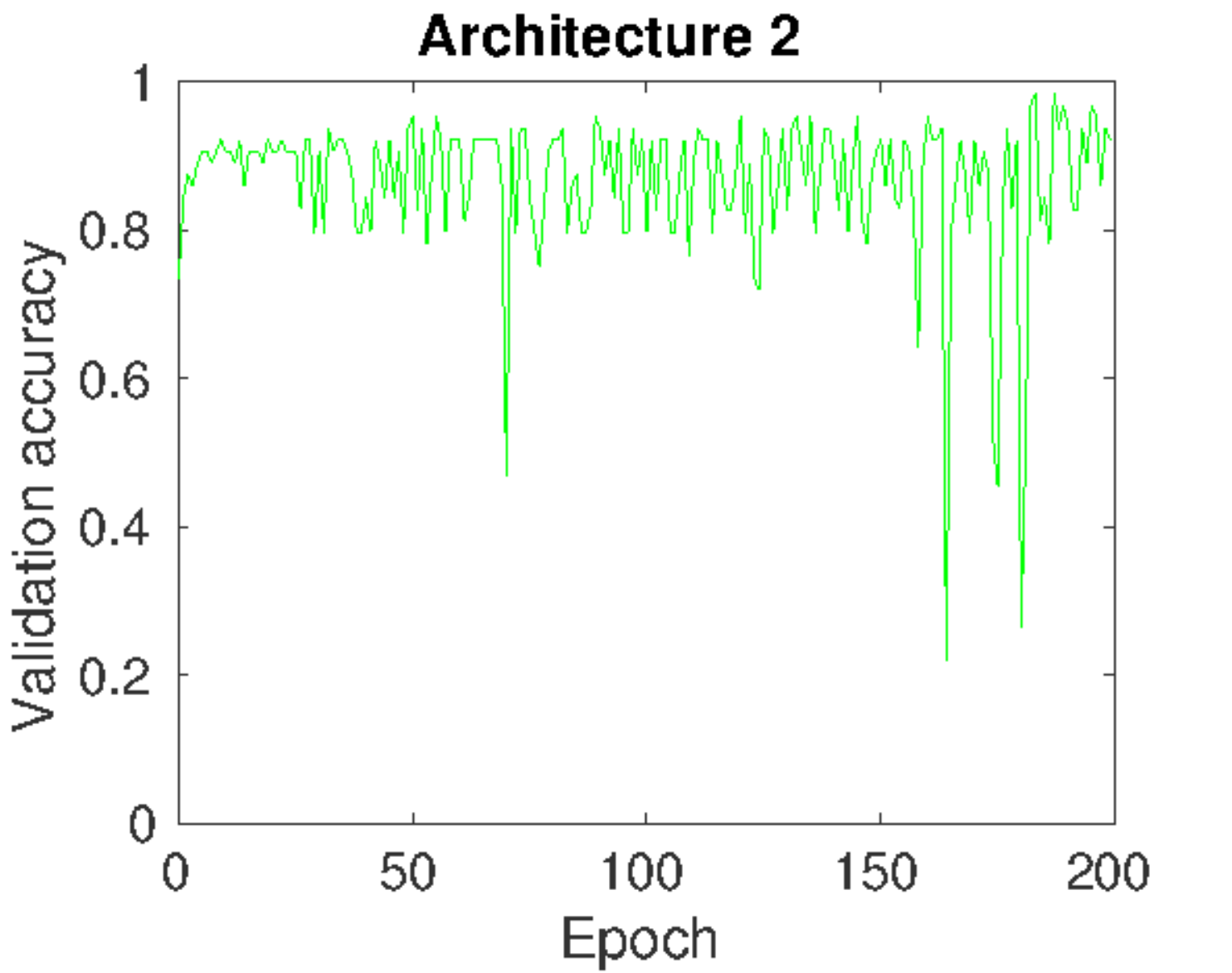}\par 
	\end{multicols}
	\begin{multicols}{2}
		\includegraphics[width=0.45\textwidth]{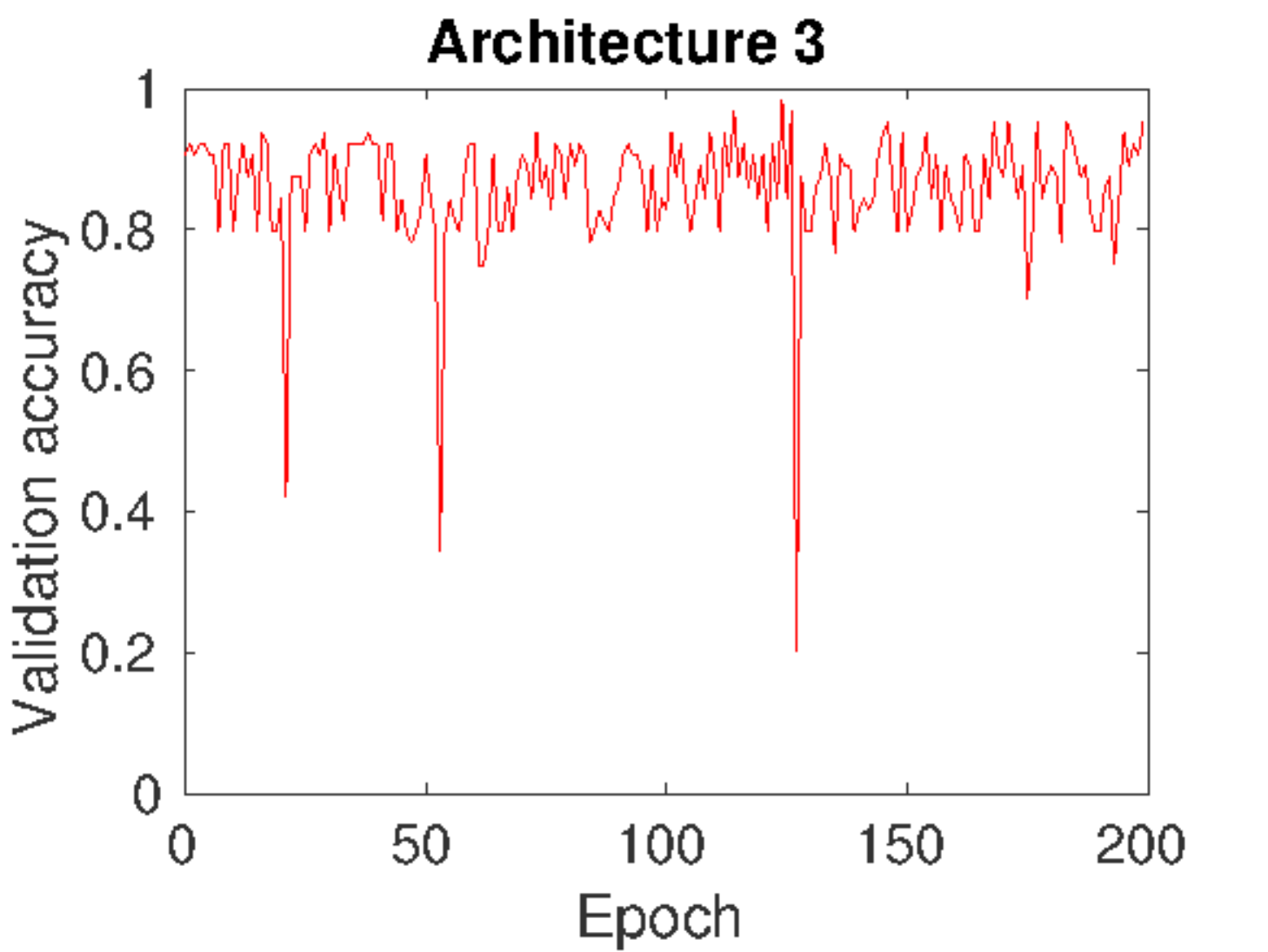}\par
		\includegraphics[width=0.45\textwidth]{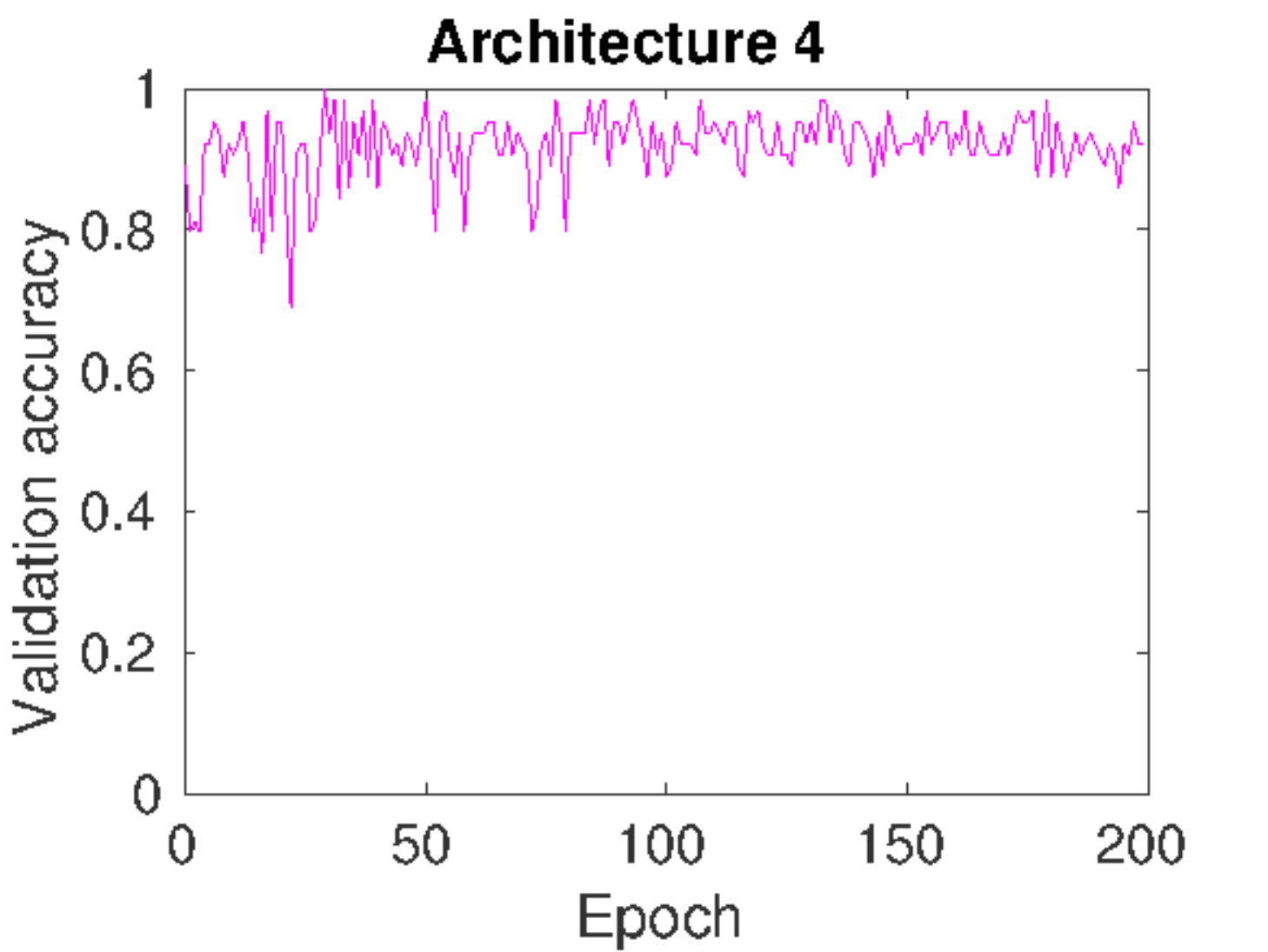}\par
	\end{multicols}
	\caption{Stability evaluation of CNN architectures. From 'Architecture 1' to 'Architecture 4', accuracy reaches stability as the number of training epochs increases. Best results were achieved with 'Architecture 4', which used a softmax layer at the top, resulting in a more stable accuracy on the validation set.} \label{archAcc}
\end{figure}

In order to choose the best model among the candidate architectures, we randomly selected 90\% of the data set for training the model, while using 10\% for validation. To deal with the great size of the data set in memory, we applied a mini-batch strategy, which consists of using several batches of $N$ images to update the final model (instead of one single block of data). After each epoch, the proposed architecture was evaluated by using the validation set. Since we focused on reducing the overfitting, the more likely architecture to be selected would be the one with high accuracy and less oscillation in the accuracy. Figure~\ref{archAcc} shows the accuracy curve for each architecture, illustrating the raise not only on the accuracy peak, but also on the stability of the curve after several epochs. Our final CNN architecture (Fig. \ref{figx:d}) consists mainly of six convolutional layers, five max-pooling layers, followed by three fully-connected and one soft-max layers for classification. The training parameters were empirically obtained through several experiments on the four architectures. The best results using \textbf{architecture 4} were achieved by training the deep network using the following parameters: 200 epochs, Adam training algorithm \citep{kingma2014adam}, $10^{-6}$ of decay rate, batch size of 32, and a learning rate of $10^{-4}$.

\begin{figure}[!t]
    \bigskip
    \bigskip
	\centering
	\includegraphics[width=0.99\textwidth]{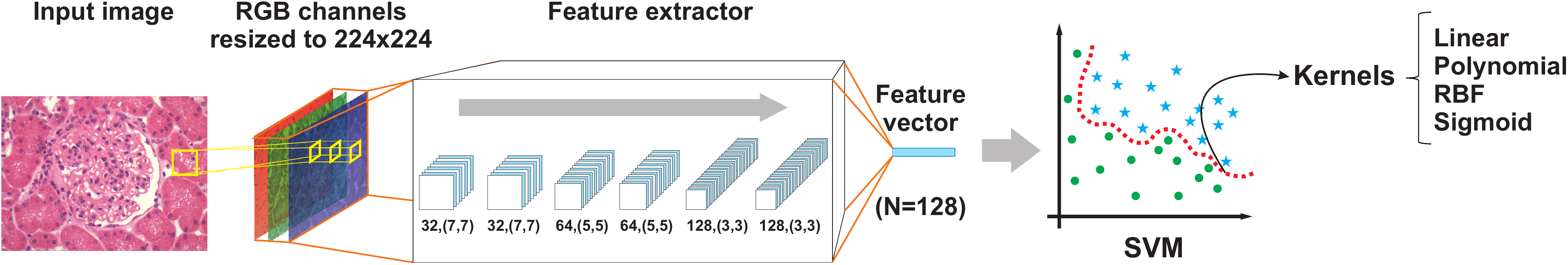}
	\caption{CNN-SVM architecture. From left to right: a glomerulus as an \textbf{input image} in an  \textbf{RGB} color space, \textbf{resized to 224$\times$224} pixels. After applying architecture 4 (Fig. \ref{figx:d}) for \textbf{feature extraction}, a \textbf{feature vector} with 128 features is generated. Finally, the resultant feature vector is classified by an \textbf{SVM} evaluated by considering \textbf{linear, polynomial, RBF and sigmoid} kernel functions.}
	\label{ypertaseis}
\end{figure}

\subsection{Classifying the CNN features with SVM}

After choosing the best architecture, the trained CNN features fed an SVM, instead of the multi-layer perceptron used for training the model. This CNN-SVM architecture was evaluated with four kernel functions: Linear, radial basis function (RBF), polynomial and sigmoid (see Fig.~\ref{ypertaseis}). 

SVM is a supervised binary classifier, which finds an optimal hyperplane to separate the classes of hypercellularity from those of normal glomeruli by using $v$ support vectors. When these classes are non linearly separable, different kernel functions can be used to map the input vectors to a higher-dimensional space (so-called \textit{feature space}), in which the input image can be linearly separated. To classify an input feature vector, SVM evaluates the sign of a function f(x), given by 

\begin{equation} \label{eq:svm}
f(x)=sign\left(\sum_{i=1}^v y_{i}\alpha_{i}N(x_{i},x_{j}) + b\right) \, , 
\end{equation}
where there are $v$-support vectors with the model parameters, $y_{i}$, and a bias parameter, $b$, laplacian coefficients from the dual optimization problem, $\alpha_{i}$. $N(x_{i}, x_{j})$ is a kernel function.

\section{Experiment analysis}

\subsection{Data set}\label{subsec:datasets}

In order to assess the performance of our proposed CNN architecture, the data set introduced in \citep{barros2017pathospotter} was used. The data set consists of 811 images, including 300 images of normal human glomerulus, while 511 images of human glomerulus with hypercellularity. As the images originated from human kidneys with different diseases, the cellular component of the hypercellularity varies among the cases. The images were selected from the digital histological image library of the Gonçalo Moniz Institute (FIOCRUZ), including cases of all the kidney biopsies performed for the diagnosis of glomerular diseases in referral nephrology services of public hospitals in Bahia state, Brazil, between 2003 and 2015. The tissue samples were fixed in Bouin’s fixative or formalin–acetic acid–alcohol, included in paraffin. Sections of 2-3 $\mu$m were stained by H\&E and PAS. The images were captured using an Olympus QColor 3 digital camera attached to a Nikon E600 optical microscope (using $\times$200 magnification). Details of the clinical and demographic characteristics of the patients from which kidney biopses were collected are presented in \citep{barros2017pathospotter}. Considering Oxford MEST, the former binary data set was relabeled into four classes: Endocapillary (90 images with endocapillary hypercellularity), with mesangial (238 images with mesangial hypercellularity), endoMes (179 images of both lesions) and normal (304 images with no lesion). In this re-evaluation process, using the MEST criteria for hypercellularity, it is noteworthy that four images were misclassified as lesioned glomeruli in the original binary data set used by \citet{barros2017pathospotter}. This occurrence led to a difference between the number of normal glomeruli on the binary corpus (300 images) and on the 4-class (304) data set.

\subsection{Methodology}

All images were resized and normalized to 224$\times$224 pixels. For a comparative evaluation considering a binary classification, a K-fold cross-validation was applied, varying K as 2, 3, 5 and 10 folds. On each iteration, 1 different fold is used for validation, and the rest ($K-1$ folds) is used for training the model. With the best CNN architecture, we compared the performance of two types of classifiers on the top of CNN backbone: CNN-MLP and CNN-SVM. Our methodology can be summarized in two steps:

\begin{itemize}
\item \textbf{CNN-MLP}: the best architecture is first found by using only 90/10 split without cross-validation. Next, using different values of K, we applied K-fold cross-validation, analyzing the performance of the models using different sizes of training and validation data.

\item \textbf{CNN-SVM}: For each value of K (folds), we selected the best CNN-MLP model. Then, we used the CNN features, obtained from the last layer before the fully-connected MLP, for the input of the SVM (see Fig. \ref{ypertaseis}).
\end{itemize}

Finally, for the multi classification, we used the same approach as the binary classification, but without varying the value of $K$. Since the 4-class data set is derived from the original data set used for binary classification, the number of images per class became smaller. This way, we decided to use $K=10$ in order to avoid a very small number of training samples per class. The one-versus-all technique was used to achieve SVM multi-class outputs.

\subsection{Evaluation metrics}\label{sec_metrics}

Four metrics were used to evaluate our proposed method: \textbf{Precision (P)} as the ratio of correctly predicting glomerular hypercellularity, and the sum of predicted true positive and false positive observations (whereby high precision is regarded to low false positive rate), \textbf{recall (R)} as the ratio of correctly predicting glomerular hypercellularity, and the sum of predicted true positive and false negative observations (whereby high recall is regarded to low false negative rate), \textbf{f1-score (F1)} as the weighted average of precision and recall (whereby high f1-score is regarded to high precision and recall rates), and, finally, \textbf{accuracy (ACC)} as the ratio of correctly predicting glomerular hypercellularity and normal glomeruli, and the total sum of positive and negative observations (whereby accuracy is proportional to true positive and true negative rates, and inversely proportional to false positive and false negative rates).

\subsection{Evaluating the proposed CNN model for binary classification}\label{best_model}

The final CNN was evaluated by using the average of the chosen metrics, observing how the model generalized the classes as the size of the training and validation set changed. It is noteworthy that a $K$ equals to 2 means a split of 50/50, as well as, K equals to 3, 5 and 10, mean 67/33, 80/20 and 90/10, respectively. Since the training set decreases proportionally to K, we used a technique of image augmentation, enlarging twice the original data set after applying pre-defined random modifications such as rotation, horizontal flip, zoom and shift. The training parameters were the same as the ones used to train the last architecture (see Section \ref{cnn_arch}). For each value of K, there were K different validation sets, resulting in K training processes and K candidate models at the ending of the training. For example, for K=10 there is one model for each training set combination, resulting in 10 models. When we evaluate only the CNN-MLP approach, the average of the metrics were computed with respect to these 10 models. However, since the aim was using the model as a feature extractor backbone, the best one out of the 10 candidates was selected, choosing the one with highest accuracy of all epochs. Table~\ref{tab5.1} shows the results of training the proposed CNN-MLP model, displaying the average metrics and their standard deviations for each train/test split. In general, all the train/test splits returned top results, achieving accuracies between 98.8\% (50/50 split) and 99.6\% (90/10 split). As expected, in the experiments using larger training sets (90/10 split), better results were achieved, although the worst scenario (50/50 split) still showed superior values for all the proposed metrics (around 98\%) in comparison with previous work \citep{barros2017pathospotter} (85\%). Another observation is the small standard deviation on all results, demonstrating the stability of the model. 

\begin{table}[t]
	\caption{Comparison between four different train/test splits with CNN-MLP on binary classification. The average metrics and their standard deviations are given for precision ($\mu$P), recall ($\mu$R), f1-score ($\mu$F1) and accuracy ($\mu$ACC).}
	\medskip
	\label{tab5.1} 
	\centering
	%\tiny
	\scriptsize
	%\footnotesize
	\begin{tabular}{c|cccc}
		\hline 
		
		\multirow{2}{*}{\textbf{Split}} & \multicolumn{4}{c}{\textbf{CNN-MLP}}   \\
		
		\cline{2-5} 
		
		\centering
		&   $\mu$P   &   $\mu$R  &  $\mu$F1   &   $\mu$Acc   \\ \hline
		
		90/10 & $\textbf{0.996} (\pm \textbf{0.009})$ &  $\textbf{0.997} (\pm \textbf{0.006})$ &  $0.995 (\pm 0.012)$ & $\textbf{0.996} (\pm \textbf{0.008})$  \\ 
		%&& \\ &&
		80/20 & $0.995 (\pm 0.008)$  & $0.994 (\pm 0.009)$  & $\textbf{0.996} (\pm \textbf{0.006)}$
		& $0.995 (\pm 0.007)$  \\ 
		
		67/33 & $0.995 (\pm 0.005)$ & $0.994 (\pm 0.005)$ & $0.995 (\pm 0.005)$
		& $0.995 (\pm 0.005)$ \\ 
		
		50/50 & $0.987 (\pm 0.003)$ & $0.987 (\pm 0.003)$  & $0.987 (\pm 0.003)$ 
		& $0.988 (\pm 0.003)$  \\ \hline
		
	\end{tabular}

\end{table}

\subsection{Choosing the best SVM kernel for binary classification}

Choosing optimal parameter values for the SVM kernel raises some questions about the interpretation of the model generated by this function and the results obtained. These questions were investigated in several works \citep{chapelle2002choosing, duan2003evaluation,imbault2004stochastic,fu2004extracting,de2006multiclass}. As shown in Table \ref{tab3}, the CNN-SVM architecture was evaluated with three parameters of kernel functions: 'C', 'gamma' and 'degree'. The regularization parameter 'C' is 1 by default, common to all SVM kernels, trading off misclassification of training examples against flatness of the solution. A low 'C' makes the classifier flatness smooth, while a high one can lead to overfitting. The 'gamma' parameter is usually 1 by default divided by number of features, and it is presented in all SVM kernels, but the linear. A small 'gamma' value represents a Gaussian distribution of the kernel function with large variance in such a way that the model might not capture the "shape" of the data set. When 'gamma' is high, the resulting model will behave similarly to a linear kernel with a set of hyperplanes separating the points of the two classes; hence, large gamma takes to high bias and low variance models, and vice-versa. The 'degree' parameter is 3 by default, and used only in polynomial kernel function. This parameter adjusts the feature space for higher-dimensional interactions. Larger 'degrees' tend to overfit the data.

\begin{table}[t]
\caption{Range of parameters to be evaluated for each SVM kernel.}
\smallskip
\label{tab3} 
\centering
\scriptsize
%\footnotesize
\begin{tabular}{c|c|c}
\hline 

 \textbf{Kernel} &  {\textbf{Function} $\mathbf{N(x_{i}, x_{j})}$} & \textbf{Parameter} \\
 
\hline
\centering
Linear & $x_{i}{ }^{T}_{ }x^{ }_{j}$ 
&  'C': [0.001, 0.01, 0.1, 1, 10, 100]   \\ \hline

RBF &   $exp (-\gamma {||x_{i} - x_{j}||^{2}})$, &  'C': [0.001, 0.01, 0.1, 1, 10, 100],\\
 &where $\gamma$ refers to gamma &
'gamma': [0.001, 0.01, 1, 1.5, 2]   \\ \hline

Polynomial &   $(\gamma$($x_{i}$${ }^{T}_{ }x^{ }_{i}) + r)^{d}$,& 'C': [0.001, 0.01, 0.1, 1, 10, 100], \\
 &where $\gamma$ denotes gamma, $r$ by coef$\theta$ & 'gamma': [0.001, 0.01, 1, 1.5, 2], \\ &and $d$ by degree &'degree':[1,2,3,4] \\ \hline

Sigmoid & $tanh(\gamma (x_{i}{ }^{T}_{ }x^{ }_{j}) + r)$, &'C': [0.001, 0.01, 0.1, 1, 10, 100],\\
 &where $\gamma$ denotes gamma &'gamma': [0.001, 0.01, 1, 1.5, 2]  \\ &and $r$ is specified by coef$\theta$ & \\ \hline

\end{tabular}
\end{table}

\begin{table}[!t]
\caption{The best results per SVM kernel on binary classification.}
\smallskip
\label{tab8} 
\centering
\scriptsize
%\footnotesize
\begin{tabular}{c|l|l|c}
\hline 

 $\textbf{Split}$ & \textbf{Kernel} & \textbf{Parameters}& \textbf{$\mu$Acc}   \\
 
\hline
\centering
\multirow{4}{*}{90/10} &  \textbf{Linear} & \textbf{'C': 1}   & \textbf{1.000} $\pm$ \textbf{(0.000)}\\ \cline{2-4}

&  \textbf{RBF} &  \textbf{C': 0.1, 'gamma': 0.001} & \textbf{1.000} $\pm$ \textbf{(0.000)}
 \\ \cline{2-4}

& \textbf{Polynomial} & \textbf{'C': 1, 'degree': 1,}&\textbf{1.000} $\pm$ \textbf{(0.000)}\\ && \textbf{'gamma': 1} & 
\\ \cline{2-4}

& \textbf{Sigmoid} &  \textbf{'C': 0.01, 'gamma': 0.01} & \textbf{1.000} $\pm$ \textbf{(0.000)}
 \\ \hline
%%%%%%%%%%%%%%%% 80_20

\multirow{4}{*}{80/20} &  Linear & 'C': 0.001   & 0.994 $\pm$ (0.011)\\ \cline{2-4}
%&& \\ &&
 &  \textbf{RBF} &  \textbf{C': 0.1, 'gamma': 0.01} & \textbf{0.996} $\pm$ \textbf{(0.010)}
 \\ \cline{2-4}

& Polynomial & 'C': 0.001, 'degree': 1,  & 0.994 $\pm$ (0.011) \\ &&'gamma': 1&
\\ \cline{2-4}

& \textbf{Sigmoid} &  \textbf{'C': 0.01, 'gamma': 0.01} & \textbf{0.996} $\pm$ \textbf{(0.010)}
 \\ \hline
%%%%%%%%%%%%%%%% 70_30

\multirow{4}{*}{67/33} &  Linear & 'C': 10   & 0.993 $\pm$ (0.006)\\ \cline{2-4}
%&& \\ &&
 &  \textbf{RBF} &  \textbf{'C': 1, 'gamma': 1} & \textbf{0.994} $\pm$ \textbf{(0.003)}
 \\ \cline{2-4}

& Polynomial & 'C': 0.001, 'degree': 3,  & 0.994 $\pm$ (0.007) \\&&'gamma': 2&
\\ \cline{2-4}

& Sigmoid &  'C': 1, 'gamma': 0.01 & 0.991 $\pm$ (0.009)
 \\ \hline

%%%%%%%%%%%%%%%% 50_50

\multirow{4}{*}{50/50} &  \textbf{Linear} & \textbf{'C': 0.01}   & \textbf{0.988} $\pm$ \textbf{(0.005)}\\ \cline{2-4}
%&& \\ &&
 &  \textbf{RBF} &  \textbf{C': 10, 'gamma': 0.01} & \textbf{0.988} $\pm$ \textbf{(0.005)}
 \\ \cline{2-4}

& \textbf{Polynomial} & \textbf{'C': 0.001, 'degree': 2,}  & \textbf{0.988} $\pm$ \textbf{(0.005)}\\&&\textbf{'gamma': 1.5} &
\\ \cline{2-4}

& \textbf{Sigmoid} &  \textbf{'C': 1, 'gamma': 0.01} & \textbf{0.988} $\pm$ \textbf{(0.005)}
 \\ \hline

\end{tabular}
%\end{table}

\bigskip

%\begin{table}[!t]
    	\caption{Comparison between four different train/test splits with CNN-SVM on binary classification. The average metrics and their standard deviation are given for precision ($\mu$P), recall ($\mu$R), f1-score ($\mu$F1) and accuracy ($\mu$ACC).}
    	\smallskip
	\label{tab5.2} 
	\centering
	%\tiny
	\scriptsize
	%\footnotesize
	\begin{tabular}{c|cccc}
		\hline 
		
		\multirow{2}{*}{\textbf{Split}} & \multicolumn{4}{c}{\textbf{CNN-SVM}}   \\
		
		\cline{2-5} 
		
		\centering
		&   $\mu$P   &   $\mu$R  &  $\mu$F1   &   $\mu$Acc    \\ \hline
		
		90/10 & $\textbf{1.000} (\pm \textbf{0.000})$ &$\textbf{1.000} (\pm \textbf{0.000})$ &$\textbf{1.000} (\pm \textbf{0.000})$ &$\textbf{1.000} (\pm \textbf{0.000})$ \\ 
		%&& \\ &&
		80/20 &$0.996 (\pm 0.006)$ & $0.996 (\pm 0.007)$&$0.996 (\pm 0.003)$	 &$0.996 (\pm 0.010)$ \\ 
		
		67/33 &$0.996 (\pm 0.004)$ &$0.996 (\pm 0.004)$ & $0.996 (\pm 0.001)$& $0.994 (\pm 0.007)$ \\ 
		
		50/50 & $0.988 (\pm 0.008)$& $0.983 (\pm 0.015)$ & $0.985 (\pm 0.004)$ & $0.988 (\pm 0.005)$  \\ \hline
		
	\end{tabular}
\end{table}

The same range of K values applied to evaluate the CNN was also used to evaluate SVM. Table \ref{tab8} shows the best parameter combinations for each kernel at each split, using accuracy as a metric for optimization. It is noteworthy that the linear kernel achieving top results means that the feature space can be linearly separable. The overall results of the CNN-SVM approach are summarized in Table \ref{tab5.2}, showing the performance of the proposed approach using the previously defined metrics. For the 90/10 split, all SVM kernels showed perfect ACC. For the 80/20 split, RBF and sigmoid kernels achieved the highest results. In the 67/33 split, RBF kernel obtained the best result. In the 50/50 split, all SVM kernels achieved the same results.

\subsection{Extending the proposed architecture to multi-classification}

\begin{table}[!t]
	\caption{Comparison between CNN-MLP and CNN-SVM models on 4-class classification with 90/10 split. The average metrics and their standard deviation are given for precision ($\mu$P), recall ($\mu$R), f1-score ($\mu$F1) and accuracy ($\mu$ACC).}
	\medskip
	\label{tab4cnn-svm} 
	\centering
	%\tiny
	\scriptsize
	%\footnotesize
	\begin{tabular}{c|cccc}
		\hline 
		
		{\textbf{Method}} &   $\mu$P   &   $\mu$R  &  $\mu$F1   &   $\mu$Acc   \\ \hline
		
		CNN-MLP & $0.925 (\pm 0.063)$ &  $0.911 (\pm 0.084)$ &  $0.913 (\pm 0.080)$ & $0.906 (\pm 0.085)$  \\ 
		%&& \\ &&
		CNN-SVM & $\textbf{0.944} (\pm \textbf{0.034})$  & $\textbf{0.945} (\pm \textbf{0.033})$  & $\textbf{0.944} (\pm \textbf{0.034})$ & $\textbf{0.945} (\pm \textbf{0.056})$  \\ \hline
		
	\end{tabular}
	%\end{table}
\medskip
%\begin{table}[!ht]
\caption{The best parameters per SVM kernel on 4-class classification.}
\medskip
\label{tab4-svm} 
\centering
%\scriptsize
\footnotesize
\begin{tabular}{l|l|c}
\hline 

 \textbf{Kernel} & \textbf{Parameters}& \textbf{$\mu$Acc}   \\
 
\hline
\centering
\textbf{Linear} & \textbf{'C': 0.01} & \textbf{0.945 $\pm$ \textbf(0.056)}\\  \hline

\textbf{RBF} &  C': 10, 'gamma': 0.001 & 0.944 $\pm$ (0.057)
 \\  \hline

\textbf{Polynomial} & \textbf{'C': 0.01, 'degree': 1} & \textbf{0.945 $\pm$ (0.056)} \\ & \textbf{'gamma': 1} & 
\\  \hline

\textbf{Sigmoid} &  \textbf{'C': 10, 'gamma': 0.001} & \textbf{0.945 $\pm$ (0.056)}
 \\ \hline

\end{tabular}
\end{table}

We also proposed the use of our CNN-SVM architecture for classification of a 4-class data set, including the following classes: Endocapillary hypercellularity, mesangial hypercellularity, \textit{endoMes} (both lesions) hypercellularity, and normal glomerulus. The same binary classification methodology was followed, but now maintaining K=10 on the cross-validation. As expected, the only modification on the CNN architecture was the number of dense layers at the top of the model, since the number of classes was changed. At each fold on cross-validation, weights from the best CNN-MLP model on binary classification were loaded, updating only the number of classes on the last layer. Then, the whole CNN-MLP model was retrained on the 4-class data set using the same former training parameters, achieving an average accuracy of 90.6\%. Just as the binary classification, the best model was selected among the 10 models from each fold of cross-validation, using the CNN backbone as a feature extractor, feeding an SVM classifier. The kernel parameters were varied in the same way as in the former experiments, achieving, as the best result, an average accuracy of 94.5\%. Table \ref{tab4cnn-svm} displays the final results for CNN-MLP and CNN-SVM classification on the 4-class data set, while Table \ref{tab4-svm} shows the parameters of the best results for each SVM kernel. The linear kernel achieved the overall best result again, proving the robustness of the CNN architecture for feature extraction. 

\section{Discussion and conclusions}

Overall on binary classification, the two classification approaches (CNN-MLP and CNN-SVM) achieved high results on all metrics with low standard deviations, as showed in Tables \ref{tab5.1} and \ref{tab5.2}. The two methods had close results, with CNN-SVM approach showing a slightly better performance for every value of $K$, proving the robustness of the final proposed model. Despite the unbalanced data set (more samples for lesion than for normal glomeruli), we did not observe the models being heavily biased on the class with more images. This behavior may be due to two factors: Image augmentation and feature quality. The process of image augmentation helped to solve this issue by increasing the number of images through random modifications on the original training set. The features obtained from the CNN backbone proved to be highly suitable for classification using all kernels, achieving an average accuracy of 100\% on the linear kernel. This outcome demonstrates that, despite the size of the CNN features 
(128), these features are linearly separable, which is an outstanding finding.

\begin{table}[!t]
\caption{Comparative performance for glomerular hypercellularity on binary classification.}
\medskip
\label{tab2} 
\centering
%\scriptsize
\footnotesize
\begin{tabular}{ccccc}
\hline 

 \textbf{Method} & \textbf{Precision} & \textbf{Recall} & \textbf{F1-Score} & \textbf{Accuracy}  \\
 
\hline
\centering
\textbf{CNN-SVM} & \textbf{1.00} & \textbf{1.00} & \textbf{1.00}  & \textbf{1.00}   \\

CNN-MLP &  $0.99$  &  $0.99$ &  $0.99$ & $0.99$  \\ 

\citet{barros2017pathospotter}  & 0.88 & 0.88 & 0.88 & 0.85 \\ \hline

\end{tabular}
\end{table}

A summary of the results of binary classification is presented in Table \ref{tab2}, displaying the best results of the CNN-MLP and CNN-SVM models in comparison with the method proposed in \citep{barros2017pathospotter}. As that previous work did not use the F1-score for evaluation, we calculated this score based on the provided precision and recall. Hence, we could compare the three results using all four metrics, considering 10-fold cross-validation (90/10 split). To the best of our knowledge, \citet{barros2017pathospotter} were the first to address the problem of glomerular hypercellularity lesion classification so far, what demonstrates that we achieved an improvement of 15 percentage points with our proposed deep learning-based model on the same data set. 

\begin{figure}[!t]
\centering
	\includegraphics[width=0.95\textwidth]{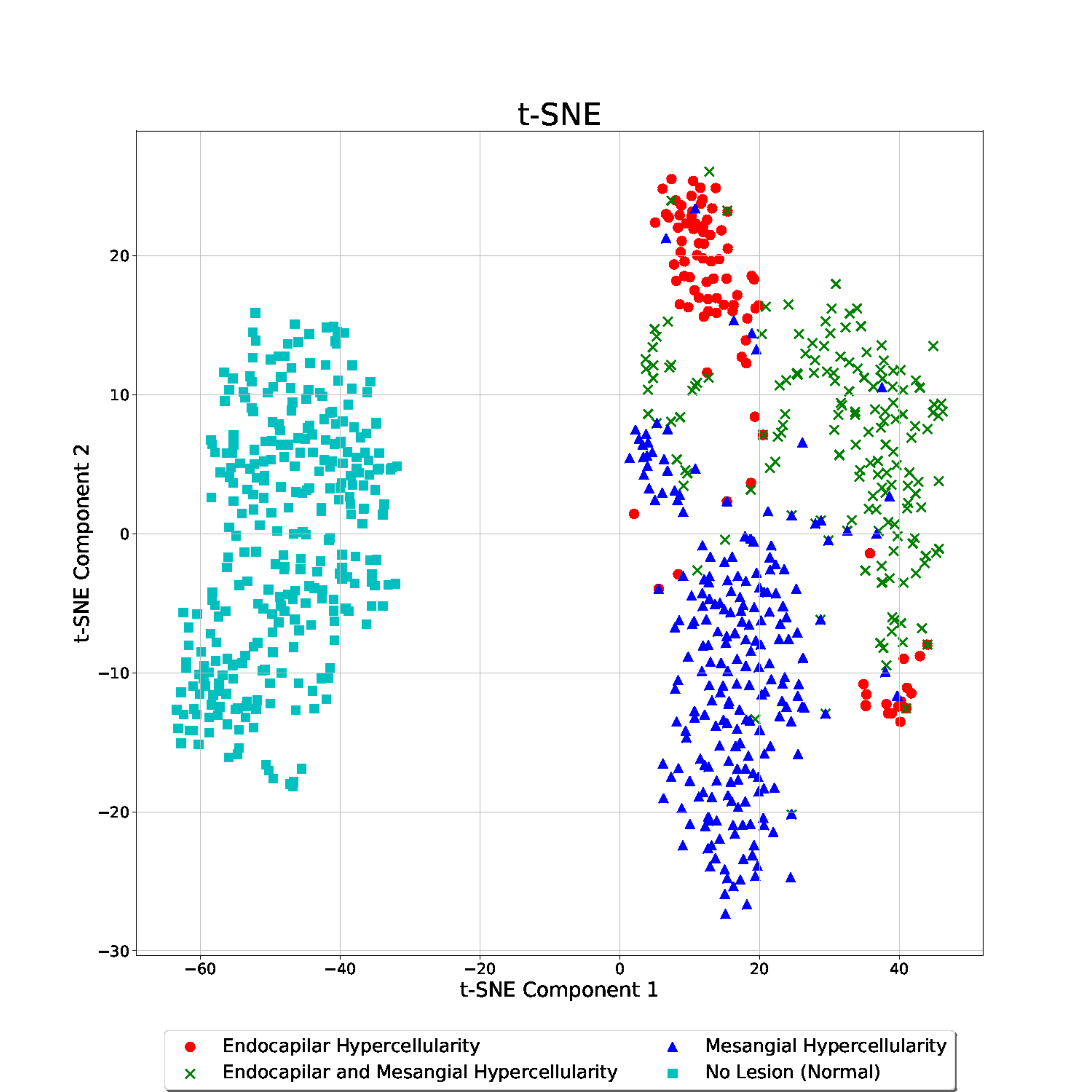}
  	\caption{t-SNE visualization of the 4-class data set. The CNN feature extractor generates a 128-dimensional feature vector, and the t-SNE algorithm reduces the dimensionality to a 2-dimensional vector to help the analysis of clusters.} \label{fig052}
  \end{figure}

Considering the 4-class classification, both CNN-MLP and CNN-SVM models achieved high results, even though the gap between these two approaches has increased (around four percentage points), as we can see in Table \ref{tab4cnn-svm}. This behavior may have occurred due to the difficulty of differentiating the 4 classes, mainly with respect to the sub-lesions. Another relevant characteristic is the \textit{endoMes} class, which contains features that can be confused with both endocapillary and mesangial hypercellularity. Figure \ref{fig052} illustrates the feature space of the data set plotted using the t-distributed stochastic neighbor embedding (t-SNE), which is a common technique for visualizing high-dimensional data into 2-dimensional plots. It's noteworthy that the "no lesion" class is well separated from the other lesion classes, which explains the 100\% accuracy of the binary classification. The three lesion classes have some well-defined groups, but these classes also have some areas with quite an overlap of instances, meaning that images containinig endocapillary, mesangial and \textit{endoMes} hypercellularity can be very similar. 

  \begin{figure}[!t]
  \centering{\subfloat[][] {\includegraphics[width=0.399\textwidth]{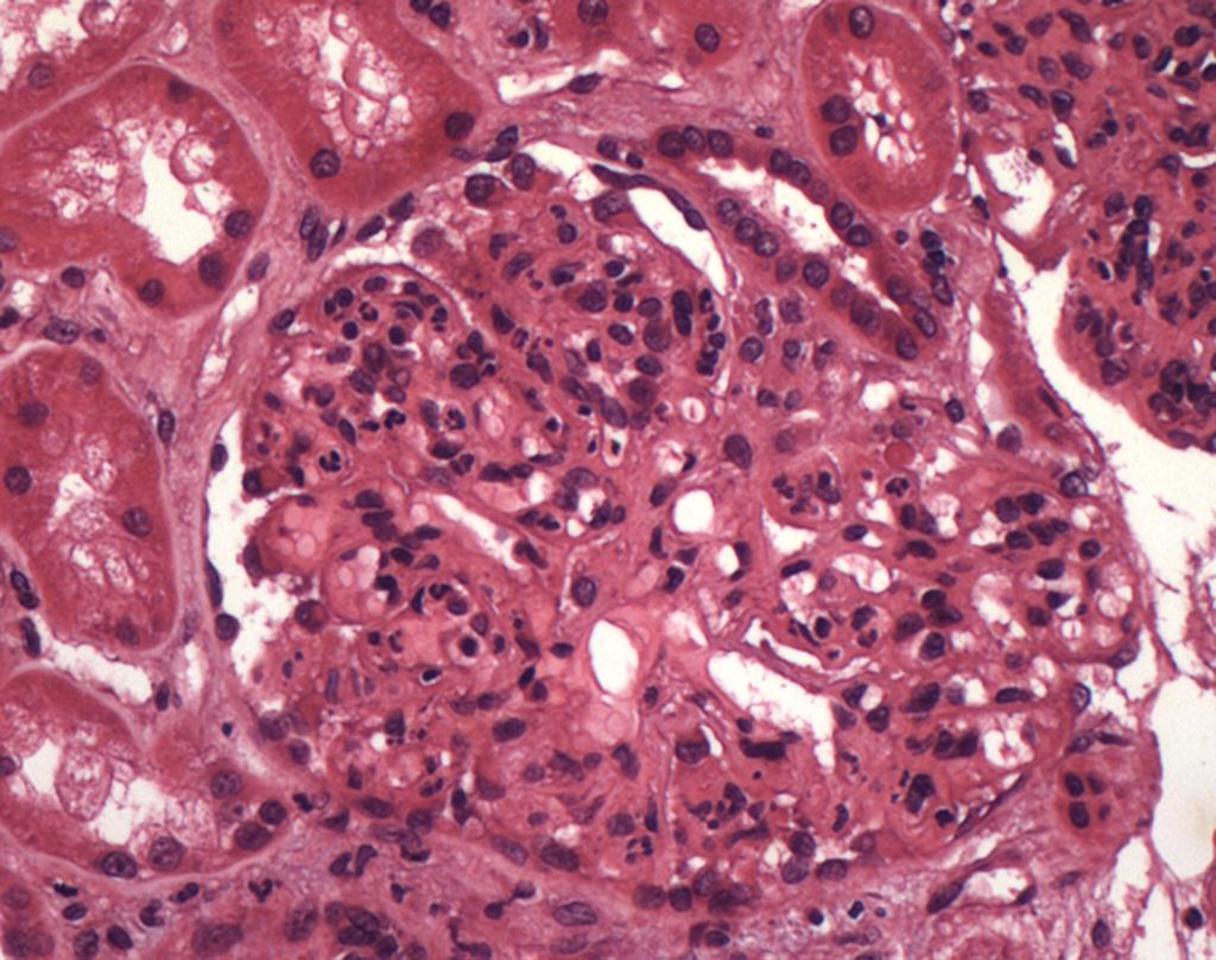} \label{figmis:a}}}
   \centering{\subfloat[][]{\includegraphics[width=0.399\textwidth]{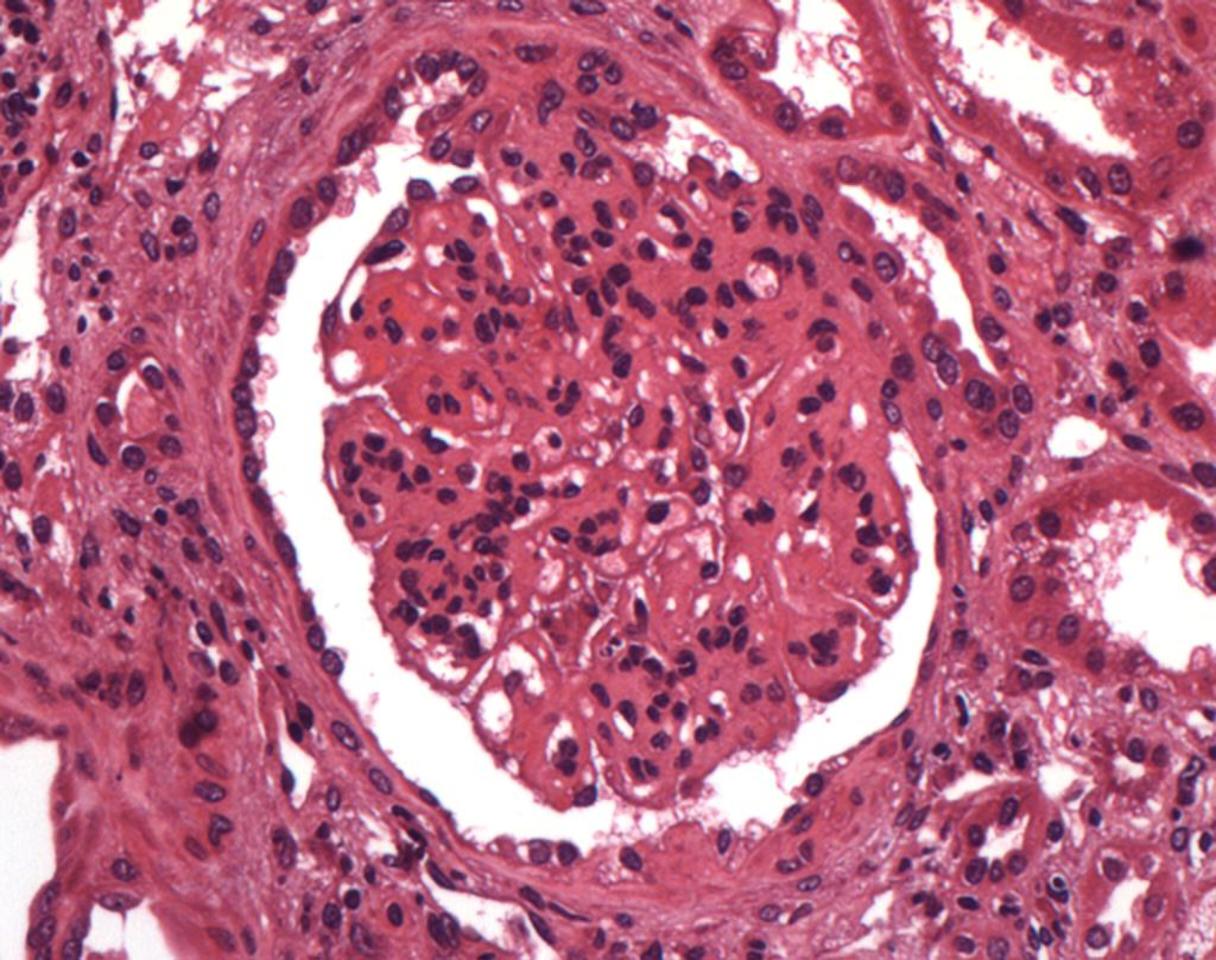} \label{figmis:b}}}
  \hfil
  \centering{\subfloat[][]{\includegraphics[width=0.399\textwidth]{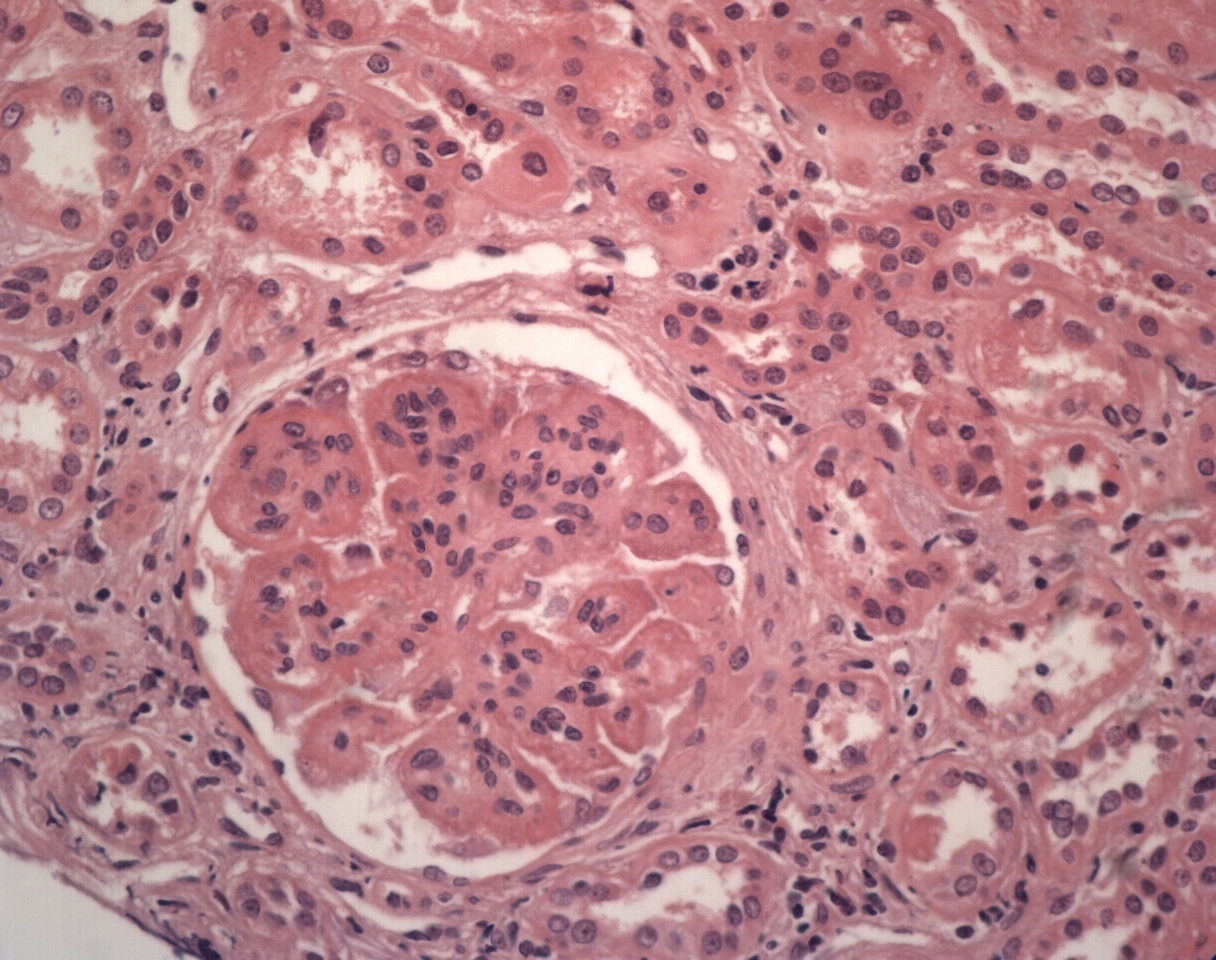} \label{figmis:c}}}
   \centering{\subfloat[][]{\includegraphics[width=0.399\textwidth]{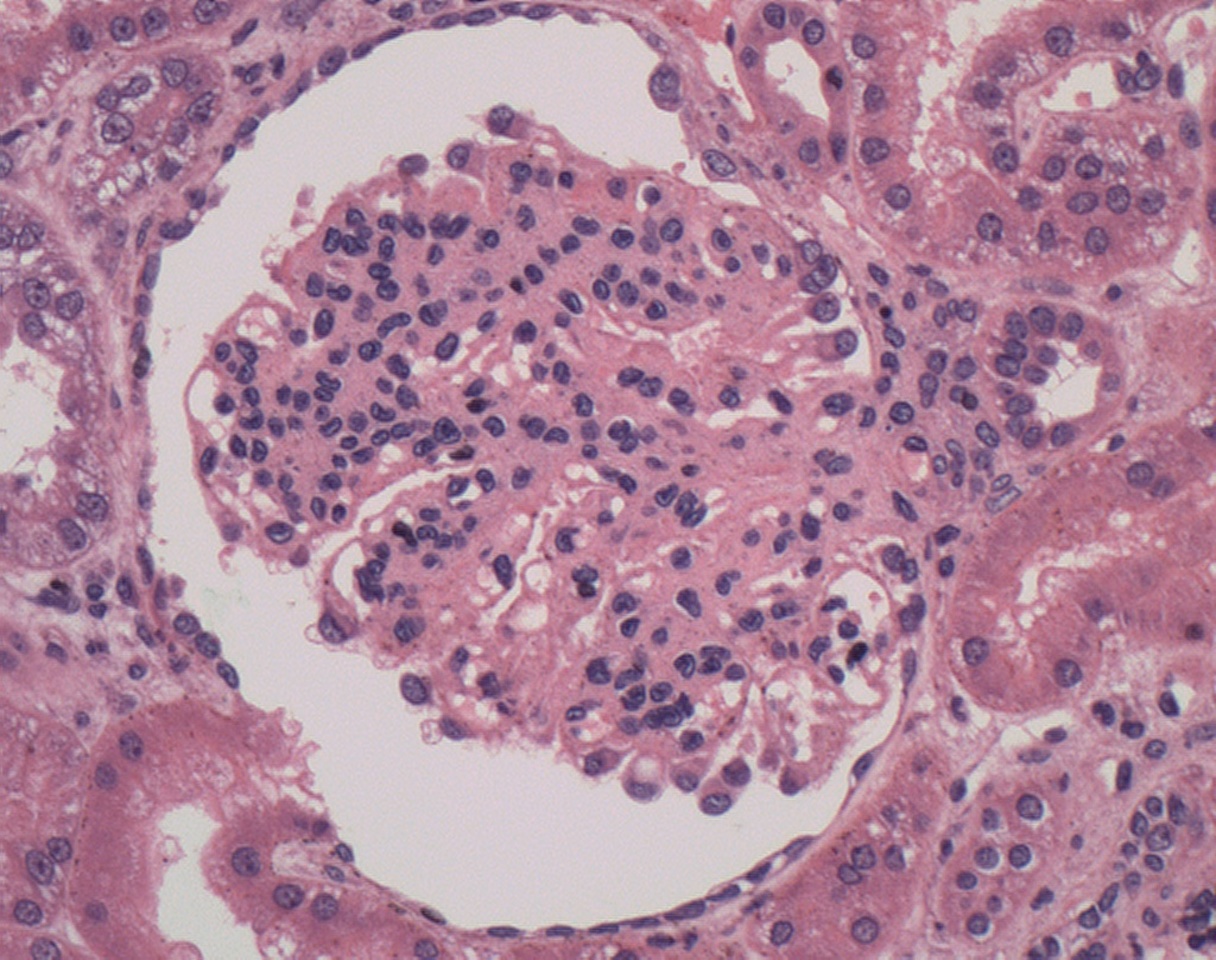} \label{figmis:d}}}
   \hfil
  \centering{\subfloat[][]{\includegraphics[width=0.399\textwidth]{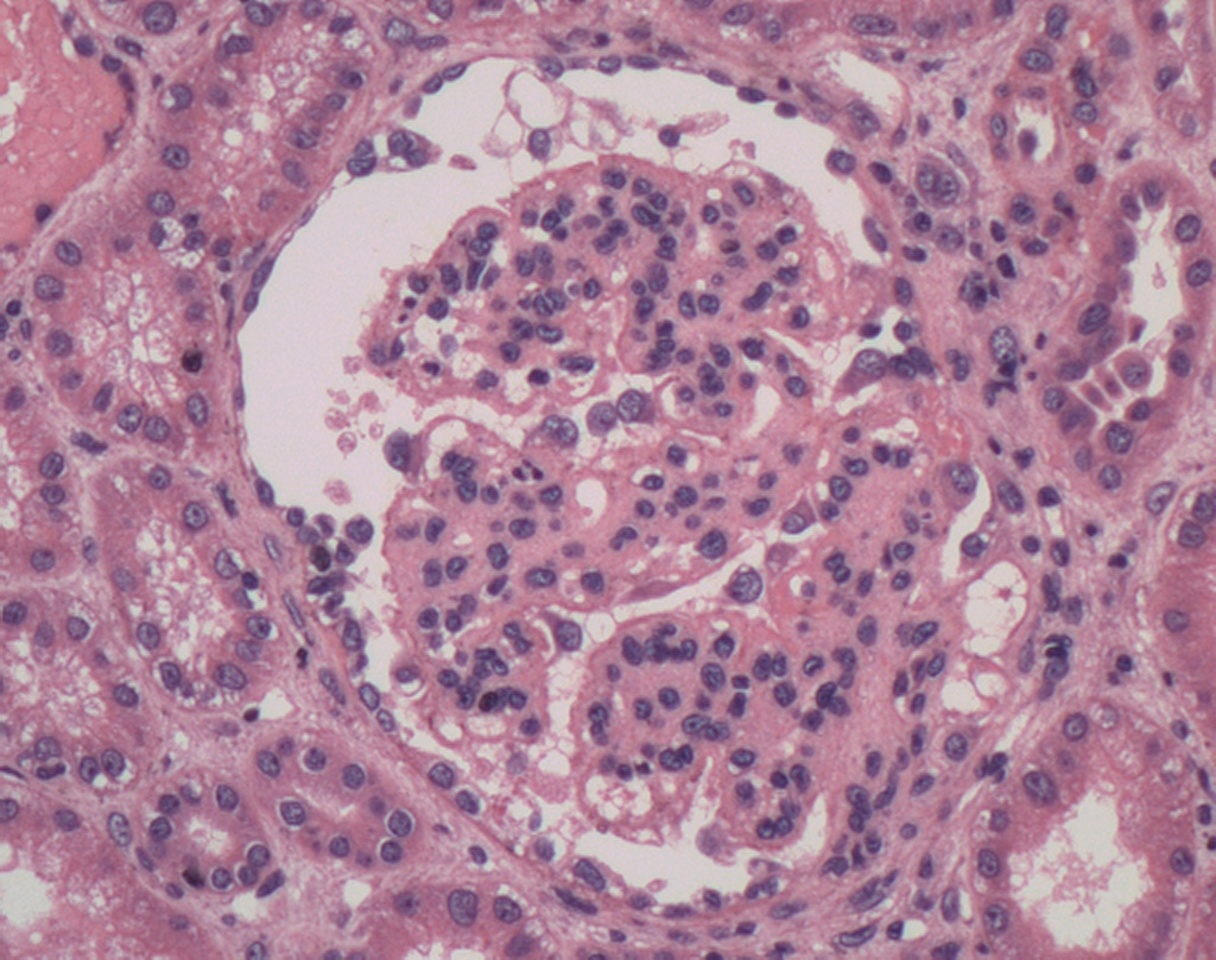} \label{figmis:e}}}
   \centering{\subfloat[][]{\includegraphics[width=0.399\textwidth]{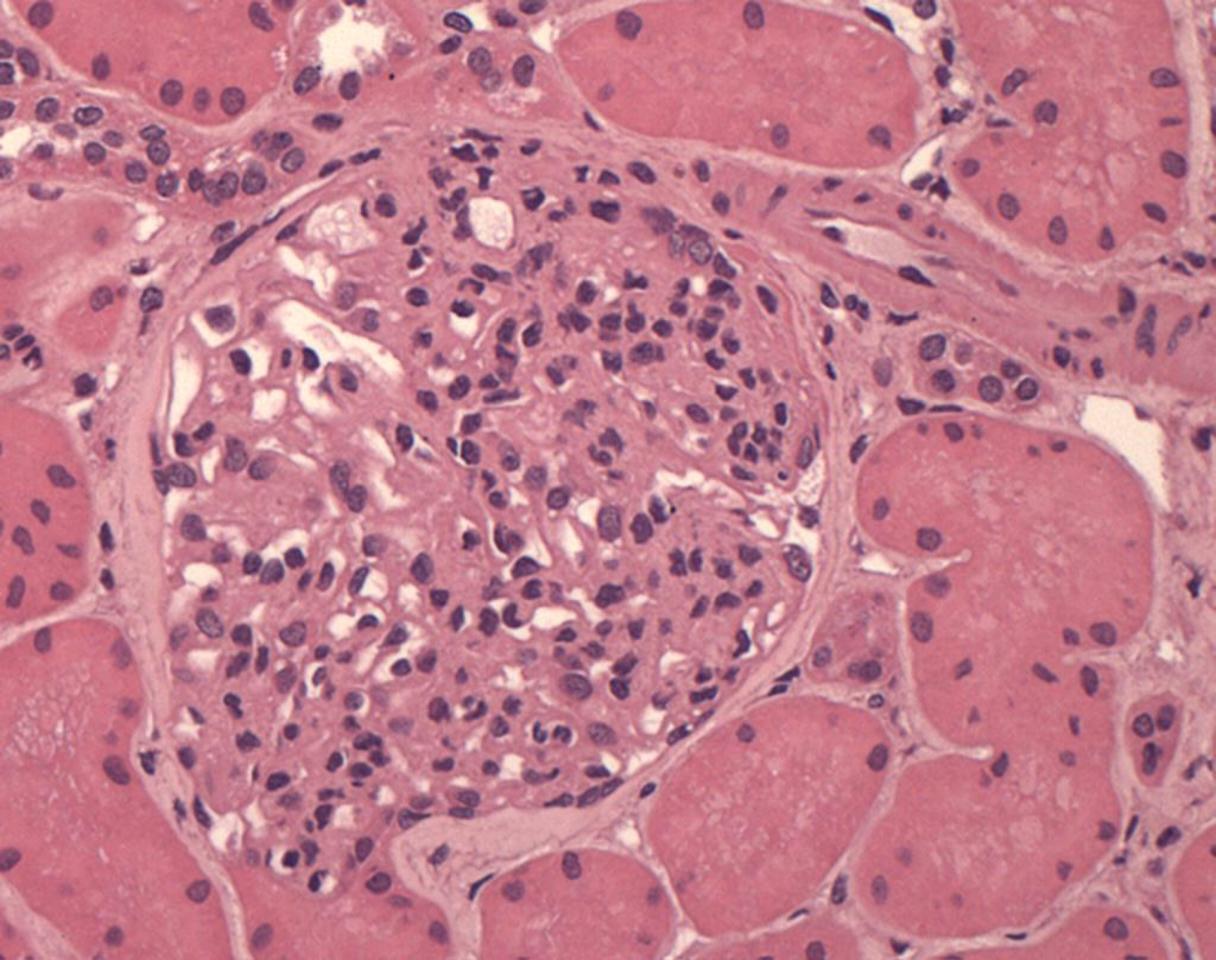} \label{figmis:f}}}
   \hfil
  
  \bigskip
 \caption{Six images of misclassified glomeruli with CNN-SVM architecture. From the left to right: (a) endocapillary hypercellularity misclassified as mesangial hypercellularity, (b) endocapillary hypercellularity misclassified as \textit{endoMes} hypercellularity, (c) mesangial hypercellularity misclassified as endocapillary hypercellularity, (d) mensangial hypercellularity misclassified as \textit{endoMes} hypercellularity, (e) \textit{endoMes} hypercellularity misclassified as endocapillary hypercellularity, and (f) \textit{endoMes} hypercellularity misclassified as mesangial hypercellularity.} \label{mis}
  \end{figure}
  
Figure \ref{mis} shows six images misclassified by the CNN-SVM model, considering every possible error combination. These images depict complex lesions that may represent a challenge even for nephropathologists (corroborating with the t-SNE visualization). Figure \ref{mis}(a) represents a glomeruli with increased circularity caused by cell proliferation and influx of inflammatory cell with disruption of glomerular compartments. Figure \ref{mis}(b) represents a glomeruli with hypercellularity combined with mesangial matrix expansion and capillary wall thickening probably by immune complex deposition on the suendothelial and on the subepithelial aspects of the glomerular basement membrane burling the limits of glomerular compartments. Figure \ref{mis}(c) hypercellularity is combined with capillary wall thickening and partial mesangial dissolution. In Figures \ref{mis}(d) and (f), mesangial and capillary lumen are not always well defined. We showed these six images to be independently classified by three pathologists. The results of this analysis are shown in Table \ref{tabpat}. Complete agreement among nephropathologists on the distribution of hypercellularity was achieved only in two out of the six images. In diagnostic practice most of the difficulties generated by these complex lesions are usually solved by examining contiguous tissue sections of 2 to 10 $\mu$m apart, stained with a variety of techniques to highlight basement membrane and mesangial matrix such as PAS and Periodic acid-methenamine silver (PAMS).

\begin{table}[!t]
	\caption{Comparison between the pathologists' labels and the results obtained by the trained CNN-SVM model. The \textit{Pool} column represents the majority voting outcome: computer (COMP) or pathologist (PAT).}
	\medskip
	\label{tabpat} 
	\centering
	%\tiny
	\scriptsize
	%\footnotesize
	\begin{tabular}{c|cccc|c}
		\hline 
		
		\multirow{1}{*}{\textbf{Image}} & \multicolumn{4}{c|}{\textbf{Classifier}} & \multirow{2}{*}{\textbf{Pool}}  \\
		
		%{\textbf{MLP}} & & & & \multicolumn{1}{|c|}{\textbf{SVM}}& & &   \\ 
		\cline{2-5} 
		
		\centering
		(see Fig.\ref{mis})&   Pathologist 1   &   Pathologist 2  &  Pathologist 3   &   CNN-SVM   \\ \hline
		
		a & END  &  END  &  END & MES &  PAT \\ 
		%&& \\ &&
		b & END  & ENDOMES  & ENDOMES & ENDOMES & COMP \\ 
		
		c & MES & END & ENDOMES	& END & COMP \\ 
		
		d & MES & MES  & MES	& ENDOMES & PAT  \\ 
		
		e & ENDOMES & MES  & ENDOMES & END  & PAT \\ 
		
		f & ENDOMES & ENDOMES  & END & MES & PAT \\ \hline
		
	\end{tabular}
	\end{table}

Although perfect results on FIOCRUZ data set have been achieved, there is a considerable gap to move from academic research to practical computational systems that assist pathologists in an effective way. For future work, we are investigating different ways of using a transfer learning approach to initialize our network with better weights for generalizing glomerulus image classes, where sufficient training data exists. Additionally, we plan to expand the number of samples (now around 31,000 unlabelled images) in the data set, working with other types of lesions and histological stains used in the pathology laboratory for better data analysis. Another work in progress is the automatic glomerulus segmentation in a WSI, containing several glomeruli; the goal is to classify each found glomerulus, considering also the individual detection of each glomerulus component.

\section{Ethical Considerations}

This work was conducted in accordance with resolution No. 466/12 of the Brazilian National Health Council. To preserve confidentiality, the images (including those shown in the paper) were separated from other patient's data. No data presented herein allows patient identification. All the procedures were approved by the Ethics Committee for Research Involving Human Subjects of the Gonçalo Moniz Institute from the Oswaldo Cruz Foundation (CPqGM/FIOCRUZ), Protocols No. 188/09 and No. 1817574.

\section*{Acknowledgment}

The work was sponsored by Funda\c{c}\~ao de Amparo \`a Pesquisa do Estado da Bahia (FAPESB) grants TO-SUS0031/2018, TO-BOL0660/2018, TO-BOL0344/2018 and TO-PET0008/2015, respectively. Washington LC dos-Santos has a research scholarship from CNPq Proc. No. 306779/2017. Luciano Oliveira has a research scholarship from CNPq Proc. No. 307550/2018-4.

\section*{References}\label{sec:reference}

%\bibliography{ref}

%\begin{comment}

%\end{comment}

\end{document}